\documentclass[aps,pre,showpacs,noshowkeys,amsmath,amssymb,amsfonts,superscriptaddress,longbibliography,reprint]{revtex4-1}
\usepackage[english]{babel}

\usepackage{graphicx}
\usepackage{bm}
\usepackage{physics}
\usepackage{mathtools}
\usepackage{gensymb}
\usepackage{tcolorbox}

\setcitestyle{super}
\usepackage{notoccite}
\usepackage{caption}
\usepackage{subcaption}
\DeclareCaptionLabelSeparator{bar}{~\rule[-0.4ex]{0.2ex}{1em}~}
\DeclareCaptionLabelFormat{subfor}{\textbf{(#2)}}
\newcommand*\bfcaption[2]{\caption[#1]{\textbf{#1.}#2}}
\usepackage{xcolor}
\definecolor{UBcolor}{HTML}{007CC1}
\usepackage[colorlinks=true,pdfnewwindow=true,linkcolor=UBcolor,citecolor=UBcolor,urlcolor=UBcolor,breaklinks=true,linktocpage]{hyperref}
\usepackage[all]{hypcap}
\usepackage[nameinlink,capitalise]{cleveref}
\begin{document}

\title{Living cells on the move}

\author{Ricard Alert}
%\email{ricard.alert@princeton.edu}
\affiliation{Lewis-Sigler Institute for Integrative Genomics, Princeton University, Princeton, NJ 08544, USA}
\affiliation{Princeton Center for Theoretical Science, Princeton University, Princeton, NJ 08544, USA}

\author{Xavier Trepat}
\affiliation{Institute for Bioengineering of Catalonia, The Barcelona Institute for Science and Technology (BIST), Barcelona, Spain, 08028}
\affiliation{Facultat de Medicina, University of Barcelona, Barcelona, Spain, 08036}
\affiliation{Instituci\'{o} Catalana de Recerca i Estudis Avan\c{c}ats (ICREA), Barcelona, Spain, 08028}
\affiliation{Centro de Investigaci\'{o}n Biom\'{e}dica en Red en Bioingenier\'{i}a, Biomateriales y Nanomedicina, Barcelona, Spain, 08028}

\date{\today}

\begin{abstract}
Spectacular collective phenomena such as jamming, turbulence, wetting and waves emerge when living cells migrate in groups.
%The quest to understand these collective behaviors stimulates the development of the physics of soft active matter.
%The quest to understand these collective behaviors stimulates the development of the physics of living matter.
\end{abstract}

\maketitle

Much like birds fly in flocks and fish swim in swarms, cells in our body move in groups. Collective cell migration enables embryonic development, wound healing and cancer cell invasion. These phenomena involve complex biochemical regulation, but their dynamics can ultimately be predicted by emerging physical principles of living matter.

\bigskip

\noindent\textbf{When and where do cells migrate in groups?}

When seen at the microscale, our body is a busy maze filled with actively moving cells. Cellular movements are slow, rarely exceeding $10$ $\mu$m/min, but crucial to embryonic development, the immune response, tissue self-renewal and wound healing. Through the same mechanisms that sustain these physiological functions, cell movements drive devastating diseases such as acute inflammation and cancer. These can in fact be considered diseases of cell movement because arresting cells in a controlled manner would be sufficient to prevent their spread. Take cancer as an example; whereas its origin is well-known to be genetic, if we could selectively stop the movement of tumor cells, we would prevent them from escaping primary tumors and reach distant organs to metastasize. Understanding cell migration is therefore crucial to improve current strategies to fight disease.

\begin{figure*}[tb]
\begin{center}
\includegraphics[width=0.75\textwidth]{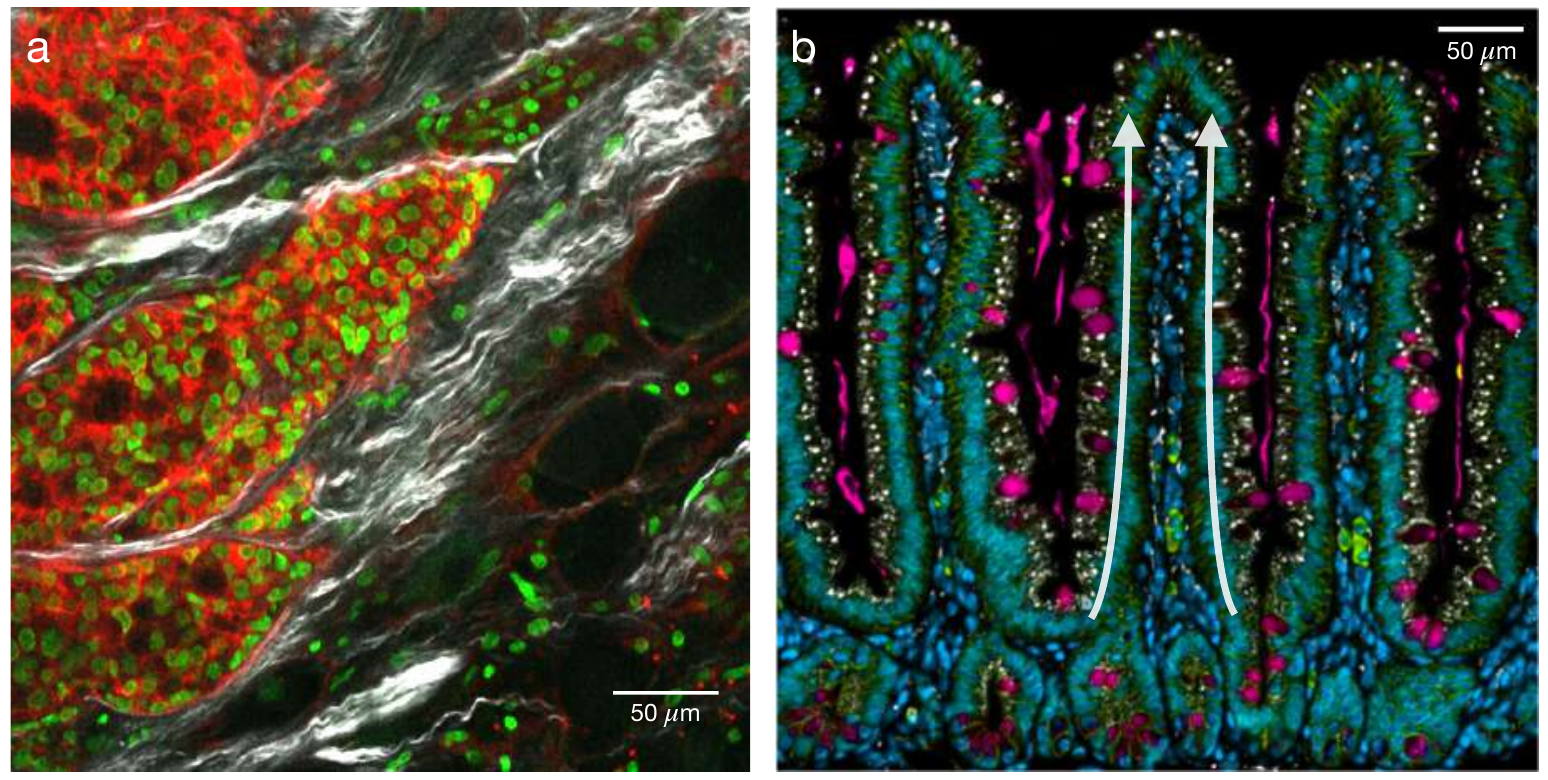}
\end{center}
  {\phantomsubcaption\label{Fig tumor-invasion}}
  {\phantomsubcaption\label{Fig intestine}}
\bfcaption{Collective cell migration in the human body}{ \subref*{Fig tumor-invasion} Breast cancer cells from a patient sample of invasive ductal carcinoma migrate collectively into the surrounding extracellular matrix. Colors label cell nuclei (green), the cell-cell adhesion protein E-cadherin (red), and collagen fibers (white). From O. Ilina et al. \textit{Dis. Model. Mech.} \textbf{11}, dmm034330, 2018. \subref*{Fig intestine} Intestine cells migrate as indicated by the white arrows to enable fast renewal of the inner surface of the gut. Colors label cell nuclei (blue), goblet cells (pink), the cell-cell adhesion protein p120 (green), and lysosomes (white). The image was kindly shared by Kristen A. Engevik. See also D. Krndija et al. \textit{Science} \textbf{365}, 705 (2019).} \label{Fig 1}
\end{figure*}

Cell migration comes in different flavors. Some cell types move as isolated self-propelled particles. For example, to chase and destroy pathogens, immune cells move individually through tissue pores. Similarly, in some types of cancer, single cells dissociate from tumors and travel solo through the surrounding tissue, eventually reaching blood vessels and metastasizing at distant organs. By contrast, in other physiological functions and pathological conditions, cells move as collectives. Collective cell movements are dominant in embryo development, where they enable the precise positioning of tissues and organ progenitors. The later coordination of growth and motion shapes organs well into postnatal life. Collective cell migration also drives wound healing in tissues such as the skin. Here, a continuous sheet of tightly adhered cells moves cohesively over the damaged area to restore a functional tissue. Cancer invasion also involves the migration of cell sheets, strands and clusters (\cref{Fig tumor-invasion}). Within these groups, cells with different function can self-organize in spatial compartments to behave like an aberrant organ. This added functionality is thought to provide malignant tumors with distinct invasive strategies that improve their chances of spreading and metastasizing. Finally, collective cell migration is also involved in maintaining the inner surface of the gut, which is the fastest self-renewing tissue in mammals. It renews entirely every 3-5 days, which implies a daily loss of tens of grams of cells. Self-renewal proceeds thanks to the division of stem cells that reside at the bottom of tissue invaginations called crypts. The progeny of these stem cells then migrates collectively from the crypt to the top of finger-like protrusions called villi, where they are shed into the intestinal lumen and discarded (\cref{Fig intestine}).

Collective cell migration is regulated by a myriad of molecular processes, from genetic programs to sensing and signaling pathways. Yet, these molecular processes act upon a limited number of physical quantities to determine cell movement. Therefore, coarse-grained physical approaches may provide crucial insight into biological questions. Furthermore, collective cell behaviors also inspire new physical theories of living systems. In this article, we highlight progress in this direction.

%\vskip0.5cm
%\begin{tcolorbox}[float=htb!]
%\textbf{The perpetual movement of cells lining the intestinal surface}

%The inner surface of the gut is the fastest self-renewing tissue in mammals. Every 3-5 days, the entire surface of the human gut is renewed, which implies a daily loss of tens of grams of cells. Self-renewal proceeds thanks to the division of stem cells that reside at the bottom of tissue invaginations called crypts. The progeny of these stem cells then migrates collectively from the crypt to the top of finger-like protrusions called villi, where they are shed into the intestinal lumen and discarded. Colors label cell nuclei (blue), goblet cells (pink), the cell-cell adhesion protein p120 (green), and lysosomes (white). The image was kindly shared by Kristen A. Engevik.

%%\begin{figure}[tbh!]
%\begin{center}
%\includegraphics[width=\columnwidth]{BoxFig}
%\end{center}
%\label{BoxFig}
%\end{figure}

%\end{tcolorbox}

\bigskip

\noindent\textbf{Cell assemblies as living matter}

What physical principles underlie collective cell motion? In traditional condensed matter, interactions between electrons and between atomic nuclei give rise to fascinating collective phenomena such as magnetism and superconductivity. Analogously, cell-cell interactions lead to emergent collective phenomena in migrating cell groups. However, when treating cell colonies as materials, we must take into account some key features of living matter, as we illustrate throughout the article.

First, the primary constituents of living tissues are cells and extracellular networks of protein fibers such as collagen. The interactions between these mesoscale constituents are orders of magnitude weaker than interatomic interactions in conventional solids. So, with notable exceptions like bone, most biological tissues are soft materials, which can easily deform and flow.

Second, cells are machines with internal engines. Specialized proteins known as molecular motors harness the energy of chemical reactions to generate forces and produce mechanical work. These energy-transducing molecular processes ultimately power cell migration, allowing cells to move autonomously, without externally applied forces. This continuous supply of energy drives living tissues out of thermodynamic equilibrium. Importantly, the driving is local; it occurs at the level of individual constituents, i.e. single cells. In other words, cells are \emph{active} constituents, and living tissues are a paradigmatic example of active matter --- an exploding new field in nonequilibrium statistical physics.

But cells are not only mechanically active: they also sense their environment, process information, and respond by adapting their behavior. For example, stem cells plated on substrates of different stiffness differentiate into distinct cell types, from brain to bone cells. So living tissues are adaptive: they respond in programmed ways to environmental cues such as external forces, the mechanical properties of the extracellular matrix, and concentrations of nutrients and signaling molecules. Consequently, cell-cell and cell-environment interactions are often very complex. Unlike atoms and electrons in conventional condensed matter, cellular interactions cannot in general be fully described via an interaction potential with a fixed functional form. Thus, a key challenge in this field is to find effective ways to capture complex cell behaviors in terms of simple, physically-motivated interactions \cite{Alert2020}.

%Some contributions to cellular interactions are mechanical, with cells acting as soft adhesive objects. These mechanical interactions can be conceivably described in terms of interaction potentials. However, other contributions to cell-cell interactions are more subtle: Cells may sense the presence of other cells via mechanical and biochemical signals.

\bigskip

\noindent\textbf{To flow or not to flow}

\begin{figure}[tb]
\begin{center}
\includegraphics[width=\columnwidth]{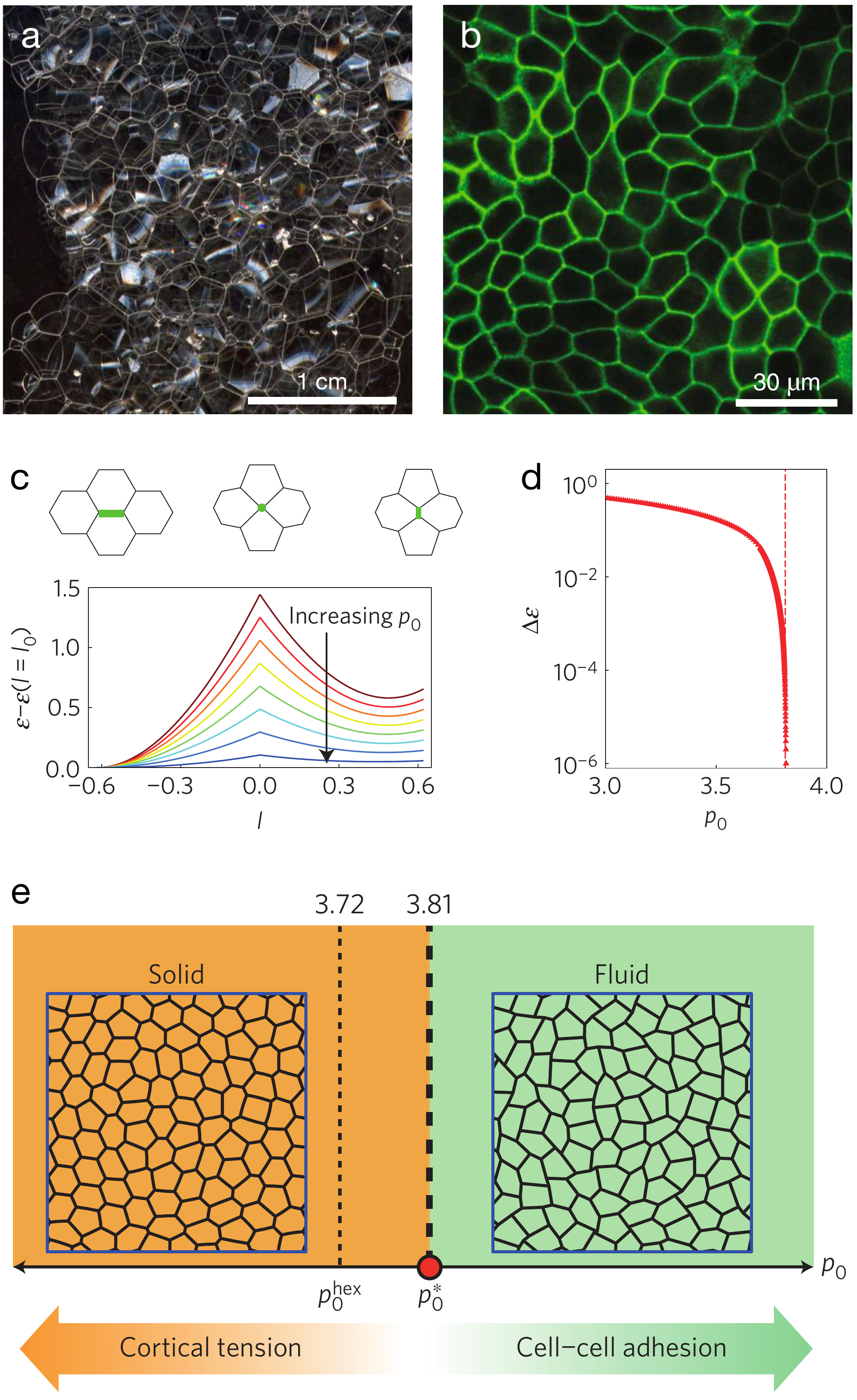}
\end{center}
  {\phantomsubcaption\label{Fig foam}}
  {\phantomsubcaption\label{Fig epithelium}}
  {\phantomsubcaption\label{Fig T1}}
  {\phantomsubcaption\label{Fig energy-barrier}}
  {\phantomsubcaption\label{Fig fluid-solid}}
\bfcaption{Biological tissues as foam}{ \subref*{Fig foam} A picture of soap foam. Adapted from Wikimedia Commons. \subref*{Fig epithelium} A monolayer of epithelial cells labelled using a membrane-targeted green fluorescent protein (green). From X. Trepat and E. Sahai. \textit{Nat. Phys.} \textbf{14}, 671 (2018). \subref*{Fig T1} For a group of four cells undergoing a T1 cell rearrangement, the energy (\cref{eq vertex}) increases as the green edge contracts ($l <0$), and it decreases as the edge expands in the perpendicular direction ($l >0$). The cell-shape parameter $p_0$ varies from 1.8 to 3.5 in equal increments. \subref*{Fig energy-barrier} The energy barrier for a T1 rearrangement decreases with $p_0$, eventually vanishing for $p_0 = p_0^* \approx 3.81$. \subref*{Fig fluid-solid} Tissues can undergo a fluid-solid transition by tuning cell shape via changes in cell-cell adhesion and/or internally-generated cortical tension. Panels \subref*{Fig T1}-\subref*{Fig fluid-solid} are from ref. \cite{Bi2015}.} \label{Fig 2}
\end{figure}

One way to think about interactions between deformable epithelial cells comes from the physics of foams. In foams, gas bubbles arrange in polygonal packings, with the liquid phase filling the interstitial spaces and providing surface tension to bubble interfaces (\cref{Fig foam}). Similarly, cells in epithelial monolayers also acquire polygonal shapes, with roughly straight edges subject to active tension generated inside cells (\cref{Fig epithelium}). This description of tissues as polygonal cell packings traces back to work by Hisao Honda and collaborators in 1980, and it was later popularized by work in the group of Frank J\"{u}licher and collaborators \cite{Farhadifar2007}. Because cells are deformable, edge lengths vary dynamically. These variations change the energy of the cellular network, which we can write in terms of areas $A_i$ and perimeters $P_i$ of cells $i=1,\ldots,N$ as
\begin{equation} \label{eq vertex}
\mathcal{H} = \sum_{i=1}^N \left[ \frac{\kappa}{2} (A_i - A_0)^2 + \frac{\Gamma}{2} (P_i - P_0)^2 \right].
\end{equation}
This energy assumes that cells resist changes in their area and perimeter around the preferred values $A_0$ and $P_0$ with elastic moduli $\kappa$ and $\Gamma$, respectively. The preferred perimeter $P_0$ depends on cell-cell interactions and cellular activity, with cell-cell adhesion promoting longer edges and cellular tension favoring shorter edges. The preferred perimeter and area define a dimensionless parameter $p_0 = P_0/\sqrt{A_0}$ which informs about the preferred cell shape. Higher $p_0$ corresponds to more elongated cells, and smaller $p_0$ corresponds to more isotropic shapes, with less perimeter for the same area.

For a given $p_0$, edge lengths vary until the system reaches its ground state, minimizing the energy in \cref{eq vertex}. In this process, an edge can shrink until it eventually disappears, leading to the formation of a new cell-cell interface (\cref{Fig T1}). These events, known as T1 transitions, allow cells to change neighbors, driving topological rearrangements of the cellular network. The ability to reorganize its constituents determines whether a material is solid or fluid. So, if cell rearrangements are difficult, the cellular network resists shear deformations; the tissue is solid. In contrast, if cells can rearrange easily, the network yields upon shear; the tissue is fluid. At small $p_0$, i.e. for roundish cells, \cref{eq vertex} reveals that there is an energy barrier preventing T1 rearrangements (\cref{Fig T1}). However, as we increase $p_0$ and cells become more elongated, the energy barrier decreases (\cref{Fig T1}). At a critical value of $p_0 = p_0^* \approx 3.81$, the barrier vanishes (\cref{Fig energy-barrier}); cells can rearrange freely \cite{Bi2015}.

This simple model thus predicts a solid-fluid transition in tissues driven by changes in cell shape, encoded in $p_0$ (\cref{Fig fluid-solid}). This is a striking prediction, which showcases the bizarre mechanical properties of materials with deformable constituents. In conventional condensed matter, solids melt by either increasing temperature or reducing the packing fraction, i.e. decreasing pressure. Tissues, however, can melt at a fixed temperature and at the maximum packing fraction, $\phi = 1$, i.e. without gaps between cells \cite{Bi2015}. Tissues can become fluid by increasing the cell perimeter-to-area ratio, for example by either decreasing intracellular tension or, counterintuitively, increasing cell-cell adhesion (\cref{Fig fluid-solid}). The more cells adhere to one another, the more they elongate, and the easier it is for them to rearrange.
%The fact that the mechanical properties of tissues depend so crucially on cell shape beautifully showcases the macroscopic consequences of having deformable constituents.

Shortly after its prediction \cite{Bi2015}, this solid-fluid transition driven by changes in cell shape was experimentally verified in layers of human bronchial epithelial cells \cite{Park2015a}. Moreover, this study showed that cells from healthy individuals tended to be caged by their neighbors and form a solid tissue, whereas cells from asthmatic individuals tended to remain unjammed, forming a fluid tissue. Therefore, these experiments suggested that the fluid-solid transition in tissues is involved in disease, opening the door to new treatments based on preventing this phase transition. Fluid-solid transitions also occur during development, enabling tissues to first turn fluid in order to remodel and acquire their shapes, and then solidify and mature. The emerging picture is that, in different biological contexts, cells can tune their shape and use the physical principles governing phase transitions in foams to decide whether to flow or not to flow.

%We can use \cref{eq vertex} to obtain the energy of the system as it undergoes a T1 rearrangement. 

\bigskip

\noindent\textbf{Aligning with neighbors}

The action starts once tissues become fluid and cells can move.
%Cell motility forces enable tissues to flow on their own, without having to apply external forces.
Cells then engage in collective flows of different types, depending on how cells align with neighbors to coordinate their individual motions. Again, these cell-cell interactions depend strongly on cell shape.

\begin{figure*}[tb!]
\begin{center}
\includegraphics[width=\textwidth]{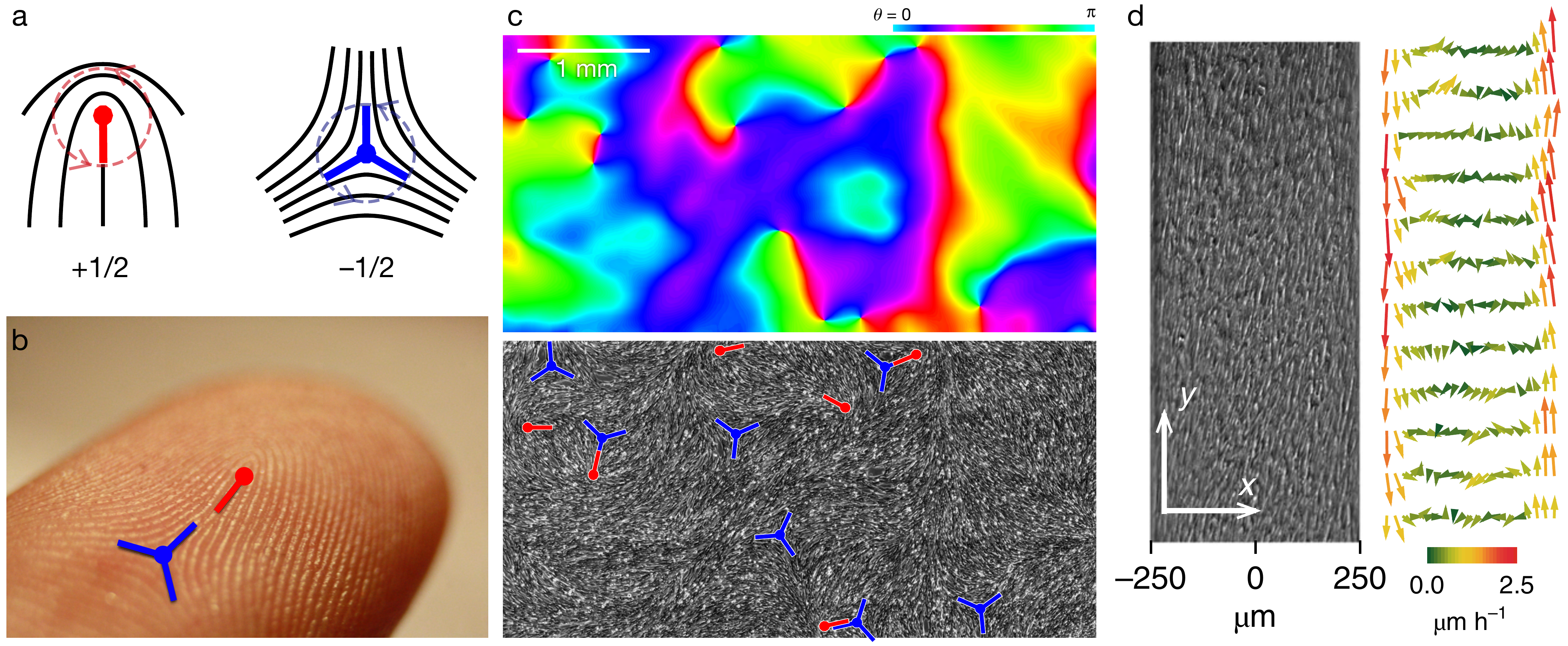}
\end{center}
  {\phantomsubcaption\label{Fig defects}}
  {\phantomsubcaption\label{Fig fingerprint}}
  {\phantomsubcaption\label{Fig cell-defects}}
  {\phantomsubcaption\label{Fig spontaneous-flow}}
\bfcaption{Topological defects and spontaneous flows in cell monolayers}{ \subref*{Fig defects} Schematics of nematic topological defects. The local alignment axis (black curves) is undefined at the defect core. Defects are characterized by their topological charge, indicated below the schematics, which is defined as the winding number of the alignment direction around the defect core. \subref*{Fig fingerprint} Topological defects in fingerprints. Adapted from Wikimedia Commons, by Frettie. \subref*{Fig cell-defects} The top panel shows the angle of cell alignment in a cell monolayer. Defects are the points where all colors meet. The bottom panel shows the defects on a phase-contrast image of the cell monolayer. Panels \subref*{Fig defects} and \subref*{Fig cell-defects} are from K. Kawaguchi, R. Kageyama, and M. Sano. \textit{Nature} \textbf{545}, 327, 2017. \subref*{Fig spontaneous-flow} A nematic monolayer of epithelial cells confined in a stripe (phase-contrast image on the left) exhibits a spontaneous shear flow (right). From ref. \cite{Duclos2018}.} \label{Fig 3}
\end{figure*}

Cells come in many different shapes: some are roughly spherical, some have rod-like shapes, and some cells develop a head-tail asymmetry to migrate persistently in one direction. This direction can be represented by a vector known as cell polarity. In groups, cells can align their individual polarities, forming phases of matter with orientational order. These alignment interactions and the resulting oriented phases can be described using concepts from the physics of magnetism and liquid crystals. For example, cells in a group can spontaneously break symmetry and align in a common direction. This type of alignment is known as polar order. To capture its emergence, one can introduce ferromagnetic-like interactions between individual cell polarities. At a coarse-grained level, collective cell polarity can be thought to result from an effective free energy with the Mexican-hat shape familiar from the Landau theory of phase transitions. This example thus illustrates how tools and concepts from other fields of physics are being borrowed to rationalize cell alignment.

In other situations, cells align along one axis but choose no preferred direction of motion.
%In this case, the alignment axis is represented not by a polarity vector but by a headless segment known as the director field.
This type of alignment is known as nematic order, taking its name from nematic liquid crystals, used in LCD screens. Prominent features of liquid crystals are singular points known as topological defects, where alignment is locally lost. You can find topological defects literally on your own hands: they are the points where your fingerprint ridges meet (\cref{Fig defects,Fig fingerprint}). Recently, researchers have discovered topological defects in several cell assemblies, from bacterial colonies to epithelial tissues, confirming that they can be described as liquid crystals (\cref{Fig cell-defects}). Interestingly, topological defects can play important biological roles. For example, in epithelial monolayers, defects promote cell death and extrusion \cite{Saw2017}. In colonies of the motile soil bacterium \textit{M. xanthus}, defects promote the formation of multicellular aggregates known as fruiting bodies, which allow the bacterial population to survive starvation \cite{Copenhagen2021}.

\bigskip

\noindent\textbf{Flowing on their own}

%\begin{figure}[htb!]
%\begin{center}
%\includegraphics[width=\columnwidth]{Fig3}
%\end{center}
%  {\phantomsubcaption\label{Fig nematic}}
%  {\phantomsubcaption\label{Fig spontaneous-flow}}
%  {\phantomsubcaption\label{Fig critical-width}}
%  {\phantomsubcaption\label{Fig turbulence}}
%\bfcaption{Spontaneous flows in cell layers}{ \subref*{Fig nematic} A nematic monolayer of epithelial cells confined in a stripe. \subref*{Fig spontaneous-flow} Spontaneous shear flow in the cell monolayer shown in \subref*{Fig nematic}. \subref*{Fig critical-width} The flow occurs only for cells in stripes wider than a critical width. Panels \subref*{Fig nematic}-\subref*{Fig critical-width} are from ref. \cite{Duclos2018}. \subref*{Fig turbulence} Normalized flow (arrows) and vorticity (color) fields showing active turbulence in large cell monolayers. From ref. \cite{Blanch-Mercader2018}.} \label{Fig 3}
%\end{figure}

Describing orientational order is not enough to account for collective cell flows. To this end, physicists describe cell assemblies as active matter \cite{Alert2020,Marchetti2013}. For example, when cells align with polar order, they can start migrating in the direction of alignment \cite{Malinverno2017}. This type of collective motion is known as flocking. The active-matter theory of this phenomenon was developed 25 years ago inspired by the mesmerizing flights of bird flocks \cite{Vicsek1995}. Today, the principles of flocking are applied to many other systems, from synthetic active colloids to bacterial swarms.

To describe nematic cell colonies, researchers use the theory of active liquid crystals, which generalizes the hydrodynamics of liquid crystals to include active (cell-generated) stresses. Among many other phenomena, this theory explains the cell flows observed around topological defects. Again, beyond cell colonies, the theory successfully describes many other systems, from biopolymer gels to shaken granular materials. Employing general theories of active matter to describe cell migration is particularly useful because it reveals connections to apparently-unrelated systems. Progressively, this approach is uncovering a classification of active systems based on symmetries and generic behaviors, in the spirit of universality classes in statistical mechanics.
%general theories of active matter describe other active systems ranging from biopolymer gels to synthetic active colloids to animal groups such as bird flocks and schools of fish. This approach reveals generic behaviors common to apparently-unrelated systems, which is progressively uncovering a classification of active systems in the spirit of universality classes in statistical mechanics.

\begin{figure*}[htb!]
\begin{center}
\includegraphics[width=0.8\textwidth]{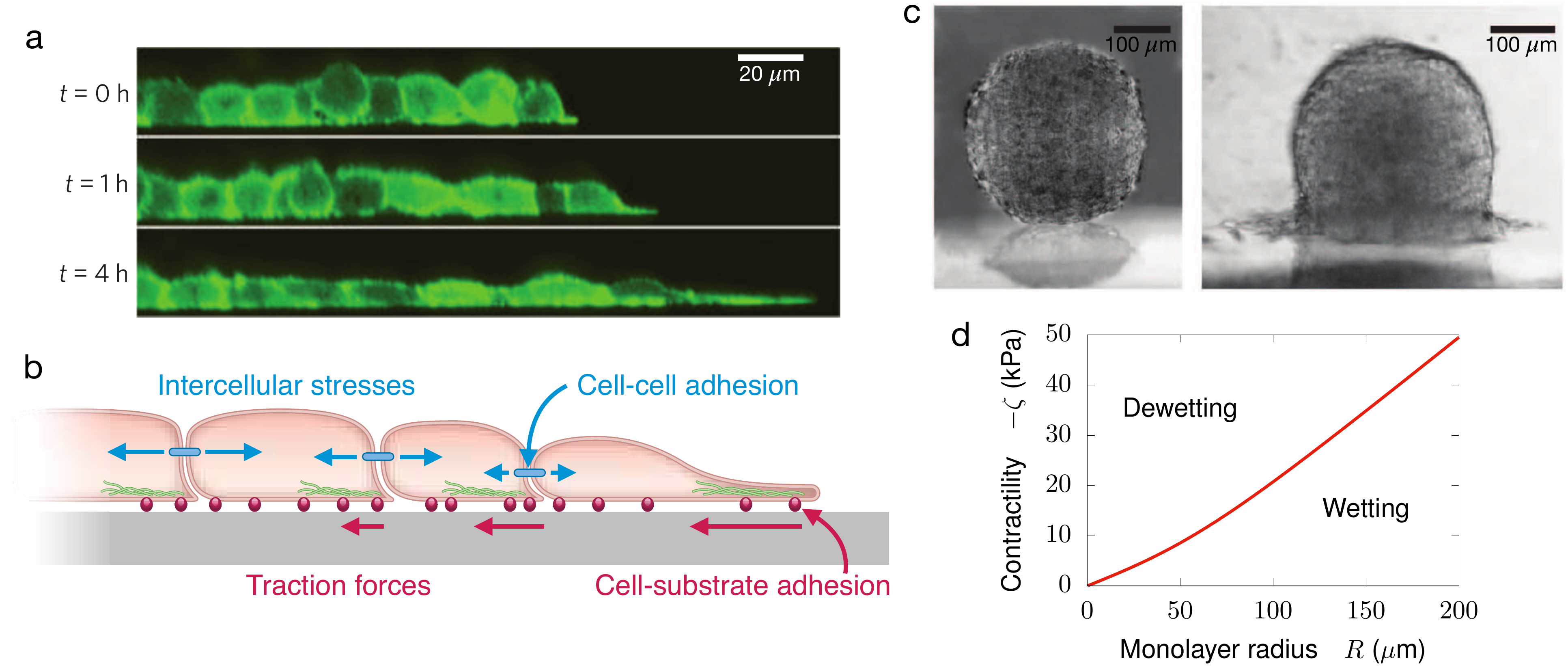}
\end{center}
  {\phantomsubcaption\label{Fig spreading}}
  {\phantomsubcaption\label{Fig forces}}
  {\phantomsubcaption\label{Fig aggregates}}
  {\phantomsubcaption\label{Fig wetting}}
\bfcaption{Tissue wetting and spreading}{ \subref*{Fig spreading} Transversal view of a cell monolayer as it starts spreading. From ref. \cite{Serra-Picamal2012}. \subref*{Fig forces} Schematic of the cell-substrate forces and cell-cell stresses involved in tissue spreading. Adapted from S.R.K. Vedula et al. \textit{Physiology} \textbf{28}, 370, 2013. \subref*{Fig aggregates} Cell aggregates may either remain as droplet-like spheroids (left) or wet the underlying substrate (right). From ref. \cite{Gonzalez-Rodriguez2012}. \subref*{Fig wetting} Phase diagram of active tissue wetting. For a given tissue contractility, which represents cell-cell pulling forces, only sufficiently large monolayers wet the substrate. Adapted from ref. \cite{Perez-Gonzalez2019}.} \label{Fig 4}
\end{figure*}

For example, the theory of active liquid crystals was originally inspired by behaviors in bacterial suspensions and in the cell cytoskeleton. Soon after formulating the theory, researchers predicted that active stresses generate an instability whereby these fluids start flowing spontaneously, without having to apply external forces \cite{Voituriez2005}.
%Rather than being driven by external forces, these spontaneous flows are powered by the internal activity.
To drive flows, active stresses have to overcome alignment forces in the liquid crystal, which happens only at sufficiently large scales.
%restoring forces that try to preserve a uniform alignment in the liquid crystal. Active stresses win this competition at sufficiently large scales, for which restoring forces are smaller.
Therefore, the theory predicted that a strip of active fluid would flow only if it was wide enough \cite{Voituriez2005}.

More than a decade later, these predictions were tested in cell monolayers \cite{Duclos2018}, showcasing the generic character of the theory. Whereas cells confined in narrow stripes did not flow, cells in stripes wider than a critical width developed a collective shear flow as predicted by the theory (\cref{Fig spontaneous-flow}). In very large tissues, cell flows become chaotic, creating disordered patterns of swirls known as active turbulence \cite{Blanch-Mercader2018}. Confinement can therefore prevent and organize spontaneous cell flows. This regulatory role of confinement may be relevant in embryonic development and tumor invasion, in which cell groups often migrate in tracks defined by the surrounding tissue (\cref{Fig tumor-invasion}). Overall, these studies showcase how cells can leverage the physics of active fluids to engage in collective flow patterns, and how confinement controls whether and how cell groups flow on their own.

%For example, polarized cells self-propel by generating a persistent traction force on the underlying substrate, much like when we walk. These active traction forces, which can be measured in experiments, have to be combined with passive forces like friction to establish the equations of motion of tissues. Following this logic, researchers have built theories for active polar media to account for the cell flows observed as polarized tissues spread into free space --- a process that mimics wound healing.

%Tissues with nematic order, in contrast, have no net polarity. Cells go back and forth.

\bigskip

\noindent\textbf{To spread or not to spread}

What happens if we release confinement and expose a cell monolayer to free space? Cells at the monolayer edge are able to sense that they have neighboring cells on one side but not on the other. In ways that remain to be fully understood, edge cells respond to this asymmetric environment by polarizing toward free space (\cref{Fig spreading}). Specifically, these cells extend protrusions known as lamellipodia, with which they exert directed and persistent traction forces on the underlying substrate to migrate toward open ground. Because cells in the monolayer are adhered to one another, the migrating edge cells pull on cells in the second row, which then also polarize, migrate, and pull on inner cells, setting the monolayer under tension \cite{Trepat2009} (\cref{Fig forces}). At the molecular level, this supracellular coordination is mediated by a protein known as merlin, which transduces intercellular forces into cell polarization. In this mechanically-coordinated way, the entire cell monolayer spreads on the substrate, becoming progressively thinner \cite{Serra-Picamal2012} (\cref{Fig spreading}). Combined with other mechanisms, this type of collective cell migration helps to close gaps in epithelial tissues, as in wound healing.

But tissues not always spread on substrates. Under certain conditions, a cell monolayer may instead retract from the substrate, eventually collapsing into a droplet-like cell aggregate \cite{Perez-Gonzalez2019} (\cref{Fig aggregates}). Tissue spreading and retraction are therefore reminiscent of the wetting and dewetting of liquid droplets. The degree of wetting depends on the balance between cohesive forces within the liquid and adhesive forces with the substrate.
%Whether a droplet spreads on a surface depends on the values of the surface tensions between the gas, liquid, and solid materials that come into contact.
By analogy, early models proposed that tissue wetting was dictated by a competition between cell-cell ($W_{\text{cc}}$) and cell-substrate ($W_{\text{cs}}$) adhesion energies \cite{Douezan2011,Gonzalez-Rodriguez2012}, encoded in a spreading parameter $S = W_{\text{cs}} - W_{\text{cc}}$.
%When $S>0$, cell-substrate adhesion dominates and the cell aggregate spreads on the substrate (wetting), whereas when $S<0$, cell-cell adhesion dominates and the aggregate retracts (dewetting).
When $S<0$, cell-cell adhesion dominates and the cell aggregate retracts from the substrate (dewetting, \cref{Fig aggregates} left), whereas when $S>0$, cell-substrate adhesion dominates and the aggregate spreads (wetting, \cref{Fig aggregates} right). This simple conceptual framework was sufficient to interpret the behavior of cell aggregates of varying cell-cell and cell-substrate adhesion \cite{Douezan2011,Gonzalez-Rodriguez2012}. However, this analogy to passive liquids does not explicitly account for the active nature of cells.
%active cell migration forces that drive tissue spreading.
%However, recent experiments showed that both tissue spreading and retraction are driven by active cellular forces \cite{Perez-Gonzalez2019}, which are not explicitly incorporated in this framework.

Recent work addressed this limitation by treating the cell monolayer as a droplet of active liquid \cite{Perez-Gonzalez2019}. Using this approach, one obtains the spreading parameter directly in terms of active cellular forces. Supported by experiments, the model predicts that the wetting transition results from the competition between two types of active forces: cell-substrate traction forces that promote spreading versus cell-cell pulling forces that promote retraction. This \emph{active wetting} framework makes another key prediction: the spreading parameter depends on the droplet radius. Cell monolayers larger than a critical radius wet the substrate, whereas smaller monolayers dewet from it (\cref{Fig wetting}). This prediction is striking because it has no counterpart in the classic wetting picture, in which the spreading parameter depends solely on surface tensions. For regular liquid droplets, size does not matter.
%; a droplet will wet a surface if surface tensions dictate so.
%In contrast, droplet-like cell aggregates may remain spherical if they are small enough but wet the surface if they overcome a critical size.
In contrast, tissue wetting is size-dependent. This prediction was verified in experiments, providing evidence for the active nature of tissue wetting \cite{Perez-Gonzalez2019}.

Besides its relevance in physics, the existence of a critical size for tissue wetting might explain drastic changes in tissue morphology during embryonic development and cancer progression. For example, a disturbing possibility is that a growing tumor might become able to spread onto the surrounding tissue once it reaches a critical size. Overall, this work exemplifies how the quest to understand collective cell migration motivates the development of new physics, leading to the discovery of phenomena like active wetting. Finally, this physics approach offers clues on how cell aggregates may tune active forces to control whether to spread or not to spread.

\bigskip

\noindent\textbf{Mechanical waves without inertia}

\begin{figure}[t!]
\begin{center}
\includegraphics[width=\columnwidth]{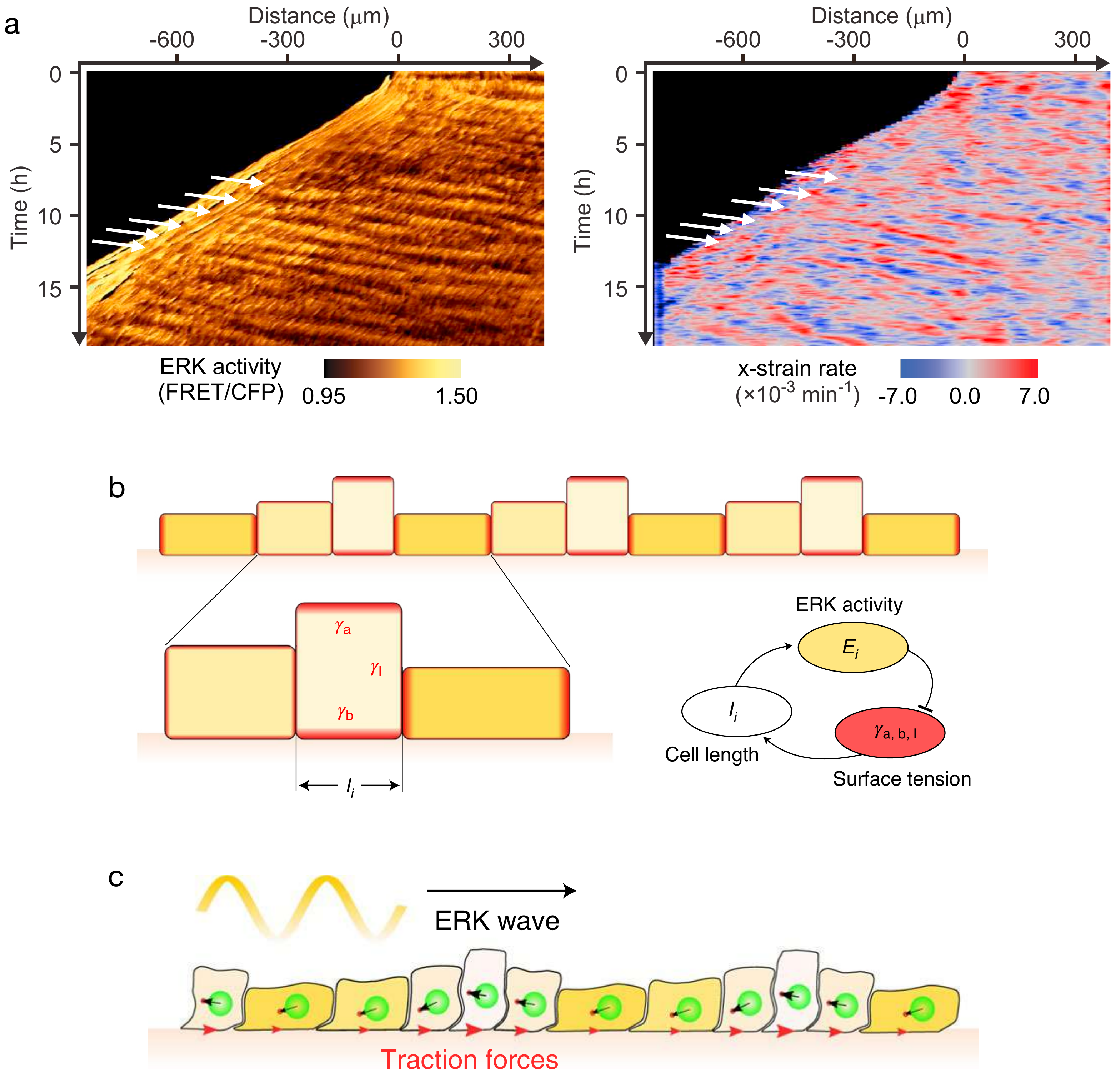}
\end{center}
  %{\phantomsubcaption\label{Fig mechano-chemical-snapshots}}
  {\phantomsubcaption\label{Fig mechano-chemical-kymographs}}
  {\phantomsubcaption\label{Fig feedback}}
  {\phantomsubcaption\label{Fig directionality}}
\bfcaption{Waves during tissue spreading}{ \subref*{Fig mechano-chemical-kymographs} Space-time plots (known as kymographs) of the activity of the signalling molecule ERK (left) and of the strain rate (right) during tissue spreading. Waves, indicated by white arrows, appear as oblique lines, whose slope gives the wave speed. From N. Hino et al. \textit{Dev. Cell} \textbf{53}, 646 (2020). \subref*{Fig feedback} Schematic of the feedback between ERK activity, cell tension, and cell length, which produces mechanochemical waves. \subref*{Fig directionality} Schematic showing that cells polarize and migrate against the mechanochemical wave. In spreading tissues, the wave travels backwards from the leading edge, directing cell migration toward free space. Panels \subref*{Fig feedback}-\subref*{Fig directionality} are from ref. \cite{Boocock2021}.} \label{Fig 5}
\end{figure}

Tissue spreading revealed yet another striking collective phenomenon: Mechanical waves start spontaneously at the leading edge and propagate across the cell monolayer \cite{Serra-Picamal2012} (\cref{Fig 5}). These waves are slow, with speeds $\sim 10-100$ $\mu$m/h, and wavelengths that span several cell diameters. Similar to longitudinal sound waves, tissue waves stretch and compress cells as they travel. Most strikingly, the waves are self-sustained; they travel long distances ($\sim 1$ mm) unattenuated. This observation is surprising because cell motion is so slow that inertia is negligible. Therefore, these waves cannot be sustained by the common back-and-forth between kinetic and potential energy familiar from the harmonic oscillator. Moreover, there are many sources of dissipation in tissues, including cell-cell and cell-substrate friction, which could potentially damp the waves. Thus, the very existence of mechanical waves in tissues implies an active driving mechanism that compensates damping and generates an effective inertia.
%In confluent, non-spreading monolayers, waves propagate isotropically, and standing waves are observed in small, confined monolayers.

The quest to understand these waves led to a plethora of physical models, ultimately revealing a palette of possible mechanisms \cite{Alert2020}. Recently, the situation has been clarified by experiments revealing that mechanical waves are accompanied by waves of ERK, an extracellular signaling molecule that affects cellular activity (\cref{Fig mechano-chemical-kymographs}). Following these observations, researchers developed a theory based on the feedback between the biochemical regulator and cell mechanics (\cref{Fig feedback}), which produces coupled chemical and mechanical waves. Assuming that cells polarize in response to stress gradients in the monolayer, the theory also explains propagation away from the leading edge in spreading tissues \cite{Boocock2021} (\cref{Fig directionality}).

These results show that cells can exploit mechanochemical feedbacks to transmit local information over long distances. This tissue-scale communication is relevant for wound healing, enabling distant cells to coordinate their migration toward the wound. Similar principles operate in morphogenesis, enabling coordinated cell deformations to precisely shape tissues without requiring local genetic control of cellular forces. From a physics perspective, this research highlights that the cells' ability to generate, sense, and respond to signals, both chemical and mechanical, can give rise to emergent phenomena as counter-intuitive as mechanical waves without inertia.

\bigskip

\noindent\textbf{Outlook}

The physics of active living matter is increasingly successful at explaining the dynamics of collective cell migration. This core biological process is being understood through concepts such as orientational order, flow, turbulence, jamming, wetting and wave propagation. The mechanistic origin and physical properties of these phenomena in cells, however, differ fundamentally from those in non-living matter. Whereas new active matter theories progressively manage to explain the broad phenomenology of collective cell migration in terms of a small number of physical variables, how cells tune these variables through thousands of genes and biochemical reactions remains a major open question.

\bigskip

\noindent\textbf{Acknowledgments}

We apologize to the many colleagues whose work could not be cited owing to space constraints. R.A. acknowledges support from the Human Frontier Science Program (LT000475/2018-C).

\bibliography{All.bib}

\onecolumngrid %this command here balances the columns before moving to the next page

\end{document}